\documentclass[11pt]{article}
\usepackage{amsmath}
\usepackage{amssymb}
\usepackage{amsfonts}
\usepackage{amscd}
\usepackage{accents}
\usepackage{pst-all}
\usepackage{epsfig}
\usepackage{graphicx,amssymb,amsmath,wasysym}
\date{\today}
\begin{document}

\begin{titlepage}

\begin{center}
{\huge \bf Challenges for Inflaton Dark Matter}
\end{center}
 
 \vspace{1cm}

 \begin{center}
{ \bf Oleg Lebedev and 
  Jong-Hyun Yoon}
\end{center}
 
  \begin{center}
 \vspace*{0.15cm}
\it{Department of Physics and Helsinki Institute of Physics, \\
Gustaf H\"allstr\"omin katu 2a, FI-00014 Helsinki, Finland
}
 \end{center}
 
 \vspace{3cm}

\begin{center} {\bf Abstract} \end{center}
 \noindent
 We examine an intriguing possibility that a single field is responsible for both inflation and dark matter,
 focussing on the minimal set--up where 
  inflation is driven by a   scalar coupling to curvature.
We study in detail the reheating process in this framework,  which amounts mainly to particle production in a quartic potential,
and distinguish thermal and non--thermal dark matter options.
 In the non--thermal case, the reheating is impeded by backreaction and rescattering,
making this possibility unrealistic. On the other hand, thermalized dark matter is viable, yet
the unitarity bound forces the inflaton mass into a narrow window close to half the Higgs mass.

\end{titlepage}

\section{Introduction}

The absence of spectacular signals of new physics in particle experiments motivates one to explore  scenarios based on minimalism.
One economical possibility   is to account for both inflation and dark matter with just a single new degree of freedom.
The corresponding Lagrangian would also be minimalistic: 
it is allowed to contain only  renormalizable interactions 
augmented with a   scalar coupling to curvature \cite{Chernikov:1968zm}, which is  in any case  induced by quantum effects.
This is a  rigid framework, yet it  may account for some of the most puzzling aspects of modern cosmology.

The discussion of possible  inflaton and dark matter unification in a more general  (often non--minimal) setting 
 started with papers by Liddle {\it et al.} \cite{Liddle:2006qz},\cite{Liddle:2008bm}. The simplest concrete model based
 on a non--minimal scalar coupling to curvature was presented in \cite{Lerner:2009xg}, where the thermal DM option
 was studied. Non--thermal inflaton dark matter in a similar  setting was recently considered in
  \cite{Tenkanen:2016twd},\cite{Almeida:2018oid},\cite{Choi:2019osi}. Other possibilities for inflaton and dark matter
  unification were explored in \cite{Khoze:2013uia}--\cite{Ji:2019gfy}.
  
 If the inflaton is stable, the Standard Model particles are produced during the inflaton oscillation epoch. 
The decay of the inflaton background  and the dynamics of the system, in general, depend crucially on collective effects.
These manifest themselves in resonant particle production \cite{Kofman:1994rk},\cite{Kofman:1997yn},\cite{Greene:1997fu}
as well as significant backreaction and rescattering \cite{Khlebnikov:1996mc},\cite{Khlebnikov:1996wr},\cite{Prokopec:1996rr}.
Perturbative estimates are often inadequate and can misrepresent the system behaviour by orders of magnitude.
 This concerns, in particular, the decay of the inflaton zero mode, the reheating temperature and other related quantities.   
 Although certain aspects of non--perturbative phenomena can be treated analytically with the help of semi--classical methods \cite{Kofman:1994rk},
 to account for backreaction and rescattering properly requires lattice simulations.
  
  In this work, we study in detail the reheating processes in the minimal inflaton dark matter model, taking into account the relevant 
  collective phenomena with the help of lattice simulations. We find that these make a crucial impact on the viability 
  of the model.

\section{Minimal  model}

The minimal inflaton dark matter model contains a real scalar  $\phi$ in addition to the Standard Model fields. The interactions of $\phi$ 
are subject to the parity symmetry 
\begin{equation}
\phi \rightarrow -\phi \;,
\end{equation}
which makes $\phi$ stable. All renormalizable interactions consistent with this symmetry are to be included. To account for inflation, one also 
includes the non--minimal scalar coupling to curvature $\phi^2 \hat R$. This interaction is expected on general grounds and is 
 induced by loop effects. The resulting inflationary potential at large field values fits the PLANCK observations very well \cite{Bezrukov:2007ep}, while
 the challenge is to understand whether $\phi$ can also fit the observed dark matter abundance. This is the main subject of the present work.

\subsection{Set--up}

The Higgs--inflaton system with non--minimal scalar--gravity couplings is described by the Lagrangian \cite{Lerner:2009xg}
 \begin{equation}
{\cal L}_{J} = \sqrt{-\hat g} \left(   -{1\over 2}  \Omega  \hat R \,  
 +  {1\over 2 } \, \partial_\mu \phi \partial^\mu \phi +  \,  (D_\mu H)^\dagger D^\mu  H  - {V(\phi,H)  }\right) \;,
\label{L-J}
\end{equation}
where $\hat g^{\mu \nu}$ is the Jordan frame metric and $\hat R$ is the corresponding scalar curvature. 
In the unitary gauge,
\begin{equation}
H(x)= {1\over \sqrt{2}} \left(
\begin{matrix}
0\\
h(x)
\end{matrix}
\right)\;,
\end{equation}
the $Z_2$--symmetric potential has the form 
\begin{equation}
V(\phi, h) = {1\over 4} \lambda_h h^4 + {1\over 4} \lambda_{\phi h }h^2 \phi^2 + {1\over 4} \lambda_\phi \phi^4  +
{1\over 2} m_h^2 h^2 + {1\over 2} m_\phi^2 \phi^2   \;.
\label{potential}
\end{equation}
We assume the mass parameters to be far below the Planck scale.
 The function $ \Omega$ includes the lowest order non--minimal scalar--gravity couplings.  In Planck units ($M_{\rm Pl}=1$),
 it reads
\begin{equation}
 \Omega = 1+  \xi_h h^2 + \xi_\phi \phi^2 \;.
 \label{Omega}
 \end{equation}
In what follows, we take $\xi_\phi , \xi_h \geq 0$ to avoid a singularity at large field values. Cosmological implications of the Higgs portal models of this
type have been reviewed in \cite{Lebedev:2021xey}.

\subsection{Singlet--driven inflation}

The scalar couplings to gravity can be eliminated by a conformal metric rescaling 
\begin{equation}
 g_{\mu\nu} = \Omega \, \hat g_{\mu\nu}   \;,
\end{equation} 
which brings us to the Einstein frame. 
In this frame, the scalar curvature term is canonical, while the kinetic terms and the potential are rescaled according to \cite{Salopek:1988qh}
 \begin{eqnarray}
 &&  K^{ij}= {3\over 2}\; {\partial \log \Omega \over \partial \phi_i} \, {\partial \log \Omega \over \partial \phi_j} + {\delta^{ij} \over \Omega}  \;, \nonumber\\
 && V_E = {V \over \Omega^2} \;,
 \label{general}
 \end{eqnarray}
 where $i,j$ label scalar fields.
In the large field limit,
\begin{equation}
   \xi_h h^2 + \xi_\phi \phi^2 \gg 1 \;,
\end{equation}
the frame function can be approximated by $ \Omega \simeq \xi_h h^2 + \xi_\phi \phi^2  $ and 
the Einstein frame Lagrangian takes the form 
 \begin{equation}
 {\cal L} = {3\over 4} \Bigl(\partial_\mu \ln ( \xi_h h^2 + \xi_\phi \phi^2 )\Bigr)^2 + {1\over 2} {1\over  \xi_h h^2 + \xi_\phi \phi^2} \Bigl(  (\partial_\mu h)^2 + (\partial_\mu \phi)^2 \Bigr)
 - {V \over ( \xi_h h^2 + \xi_\phi \phi^2)^2} \;.
 \end{equation} 
 Introducing the variables \cite{Lebedev:2011aq}
 \begin{eqnarray}
 && \chi= \sqrt{3\over 2} \ln ( \xi_h h^2 + \xi_\phi \phi^2) \;, \nonumber\\
 && \tau = {h \over \phi} \;,
 \end{eqnarray}
 one may rewrite the kinetic terms  as
 \begin{eqnarray}
   {\cal L}_{\rm kin} &=& {1\over 2} \left(    1+{1\over 6} { \tau^2 +1 \over \xi_h \tau^2 + \xi_\phi}   \right) \, (\partial_\mu \chi)^2 +
 {1\over \sqrt{6}}  {(\xi_\phi -\xi_h)\tau \over  ( \xi_h \tau^2 + \xi_\phi)^2 }\,  (\partial_\mu \chi)(\partial^\mu \tau)  \nonumber\\
 &+&    {1\over 2} {  \xi_h^2 \tau^2 + \xi_\phi^2 \over (   \xi_h \tau^2 + \xi_\phi )^3   } \, (\partial_\mu \tau)^2  \;.
 \label{full-kinetic}
 \end{eqnarray}
 In general, $\tau$ and $\chi$ mix, while if the (local) minimum of the potential is  at
\begin{equation}
\tau_{\rm min} =0 \;,
\end{equation}
 the mixing vanishes. Inspecting the Einstein frame potential at large $\chi$,
 \begin{equation}
V_E = { \lambda_h \tau^4 + \lambda_{\phi h } \tau^2 +\lambda_\phi  \over 4 ( \xi_h \tau^2 + \xi_\phi)^2 } \;,
\end{equation}
one finds that  $\tau=0$ is a local minimum if
 \begin{equation}
   \lambda_{\phi h} \xi_\phi - 2 \lambda_\phi \xi_h >0 \;.
          \end{equation}
The canonically normalized fields at this point are
\begin{equation}
    \chi^\prime = {\chi \; \sqrt{1+{1\over 6\xi_\phi} }}~~,~~ \tau^\prime={\tau \over \sqrt{\xi_\phi}} \;.  
       \end{equation}
At leading order in $1/( \xi_h h^2 + \xi_\phi \phi^2)$, the potential is flat with respect to $\chi$: $V_E = \lambda_\phi / (4 \xi_\phi^2)$.
The $\tau^\prime$ field behaves as a heavy spectator if $m_{\tau^\prime} \gtrsim  H$, where $H$ is the inflationary Hubble rate. 
Computing $m_{\tau^\prime}$ from the above potential, one finds that 
$\tau^\prime$ can be integrated out when
\begin{equation}
   \lambda_{\phi h} \gtrsim \lambda_\phi \; {12 \xi_h +1 \over 6 \xi_\phi} \;.
   \label{constraint-xi}
             \end{equation}
If this inequality is violated, the dynamics of the system do not reduce to that of a single field $\chi^\prime$. We note that the choice $\xi_h=0$ is allowed 
by this constraint, if $ \lambda_{h\phi}$ is sufficiently large.

Suppose that the constraint (\ref{constraint-xi}) is satisfied. Since the potential is almost flat in the $\chi^\prime$ direction, it is a natural inflaton 
candidate. Retaining the next to leading term in the $1/( \xi_h h^2 + \xi_\phi \phi^2)$ expansion of (\ref{general}), one finds
 \begin{equation}
V_E = {\lambda_{\phi} \over 4 \xi_\phi^2} \left( 1+  \exp \left(     - {2 \gamma \chi^\prime \over \sqrt{6}}  \right) \right)^{-2} \;,
  \label{VE1}
 \end{equation}
 where 
  \begin{equation}
  \gamma = \sqrt{6 \xi_\phi \over 6\xi_\phi +1  } \;.
 \end{equation}
There is no contribution to the potential from the $\tau-\chi$ mixing at this order since $\partial V_E /\partial \tau =0$ at the minimum.
The above potential is the same as that for Higgs inflation \cite{Bezrukov:2007ep} which requires $\gamma \simeq 1$, while in our case $\gamma $ can be below 1.

The inflationary predictions are read off from the slow roll parameters, 
 \begin{eqnarray}
&&    \epsilon = {1\over 2} \left( {\partial V_E / \partial \chi^\prime \over V_E }    \right)^2    \;, \nonumber\\
&& \eta = {\partial^2 V_E / \partial \chi^{\prime\,2 } \over V_E }   \;.
      \end{eqnarray}
Inflation ends when $\epsilon \simeq 1$, which determines $\chi^\prime_{\rm end}$. The number of $e$--folds is given by
 \begin{equation}
N= \int_{\rm in}^{\rm end } H \, dt = - \int^{\rm end }_{\rm in} {V_E \over \partial V_E / \partial \chi }\, d \chi \;.
 \end{equation}
For a given $N$, this equation defines the initial $\chi^\prime_{\rm in}$. The COBE constraint on inflationary perturbations requires $V_E/ \epsilon \simeq 0.027^4$ at 
$\chi^\prime_{\rm in}$, which implies
\begin{equation}
 {\lambda_{\phi} \over 4 \xi_\phi^2} = 4\times 10^{-7} \, {1\over \gamma^2 N^2 } \;.
 \label{cobe}
 \end{equation}
The spectral index $n$ and the tensor to scalar ratio $r$ are computed via
   \begin{eqnarray}
&&     n = 1-6 \epsilon + 2 \eta \simeq 1-{2\over N}  - {9\over 2 \gamma^2 N^2 }  \;, \nonumber\\
&&  r =16 \epsilon \simeq {12\over \gamma^2 N^2}  \;.
\label{n-r}
      \end{eqnarray} 
These predictions fit very well the PLANCK data for $N =50$ to 60 \cite{Akrami:2018odb}.

It is important to note that $\gamma \ll 1$ is inconsistent with our approximations, i.e. the expansion in $1/(\xi_h h^2 + \xi_\phi \phi^2)= 
 \exp \left(     - {2 \gamma \chi^\prime / \sqrt{6}}  \right)$. For $\gamma \lesssim 1/\sqrt{N}$, inflation takes place in the regime 
$ \exp \left(     - {2 \gamma \chi^\prime / \sqrt{6}}  \right) \gtrsim 1$, where higher order terms  are important.

Further constraints on the model are imposed by unitarity considerations. A non--minimal coupling to gravity corresponds to a non--renormalizable dim--5
operator, which implies that the model is meaningful up to the unitarity cutoff, namely, $ \Lambda \sim 1/\xi_\phi$ in Planck units \cite{Burgess:2009ea},\cite{Barbon:2009ya}. 
 In particular, the inflationary scale
$(\lambda_\phi/ 4\xi_\phi^2)^{1/4}$ must be below the cutoff. It should be noted that  $\Lambda$ depends on the inflaton background value \cite{Bezrukov:2010jz},
however at  the end of inflation and beginning of reheating, this background  becomes insignificant, such that the cutoff is around $1/\xi_\phi$ (see also \cite{Ema:2016dny}).
 Combined with (\ref{cobe}),    this requires at the inflationary scale \cite{Ema:2017ckf}
\begin{equation}
 \lambda_{\phi} (H)< 4\times 10^{-5} 
 \label{uni-bound}
 \end{equation}
and $\xi_\phi (H) <300$. Here we have set    $\gamma \sim 1$ since the bound is only relevant for large $\xi_\phi$.

\subsection{Non--thermal inflaton dark matter and reheating}

The critical question in this model is how the inflaton energy gets converted into SM radiation. Since the direct inflaton decay is forbidden, this energy transfer can only happen
during the inflaton oscillation epoch. After that, the total number of the inflaton quanta remains constant due to its feeble interactions.

Since the presence of a non--trivial $\xi_h$ is  inessential for our discussion, 
let us now focus on the case $\xi_h =0 $   and $   \lambda_{\phi h} \gtrsim \lambda_\phi / (6 \xi_\phi)$, which implies that  the Higgs is a heavy spectator  stabilized at $h\simeq 0$  during inflation.
At the end of  inflation, the inflaton field value is around the Planck scale. Therefore,  the Planck mass cannot be neglected in $\Omega$  leading to a complicated dependence of the canonically normalized inflaton $\chi$ on $\phi$,
 \begin{equation}
{d \chi \over d \phi} = \sqrt{  1 + \xi_\phi (1+6 \xi_\phi) \phi^2  \over  (1 +\xi_\phi \phi^2)^2 } \;.
\label{dchi/dphi}
 \end{equation}
 This equation is solved by \cite{GarciaBellido:2008ab}
  \begin{equation}
  \, \chi (\phi) = \sqrt{ 1+6 \xi_\phi \over \xi_\phi} \, \sinh^{-1} \left(    \sqrt{(1+6 \xi_\phi) \xi_\phi} \, \phi  \right)
 -\sqrt{6 } \,  \sinh^{-1} \left(       {\sqrt{6} \xi_\phi    \phi \over \sqrt{1+ \xi_\phi \phi^2}}            \right)\;.
  \end{equation}
Shortly after inflation, the inflaton amplitude decreases such that $\xi_\phi \phi^2, \xi^2_\phi \phi^2 <1 $ and
\begin{equation}
\chi \simeq \phi \;.
\end{equation} 
Therefore, the inflaton oscillates in the quartic potential,\footnote{We assume the bare inflaton mass to be far below the Planck scale so that it can be neglected during (p)reheating.
At large $\xi_\phi$, however, there is  a brief period in which the inflaton potential is quadratic: 
$V_E \simeq \lambda_\phi \chi^2 / (6 \xi_\phi^2)$ for the field range $1/(2\xi_\phi) \ll |\chi | \ll 1$.  We find that this feature is unimportant for our purposes (see also \cite{Ema:2017ckf}).}
\begin{equation}
V_E (\phi) \simeq {1\over 4 } \lambda_\phi \,\phi^4\;.
\end{equation}
At this stage, the Higgs quanta start getting  produced by a time dependent inflaton background. Since the fields are effectively massless, particle production takes place in a 
quasi--conformal regime, which means that all the relevant quantities redshift in the same fashion leading to strong Bose--Einstein enhancement of the process \cite{Greene:1997fu}.

Let us consider the main features of particle production  in a $\phi^4$ potential \cite{Greene:1997fu}. 
The classical inflaton background obeys the equation of motion
 \begin{equation}
 \ddot\phi + 3H \dot \phi + \lambda_\phi\, \phi^3 =0 \;.
 \end{equation}
After a few oscillations, the solution takes the form of a Jacobi cosine,
 \begin{equation}
\phi(t) = {\Phi_0 \over a(t)} \; {\rm cn} \left( x, {1\over \sqrt{2}} \right) \;,
 \end{equation}
 where $x=(48 \lambda_\phi)^{1/4} \sqrt{t} $ is the conformal time. The scale factor is fixed by  $a(0)=1$ and $a(t) \propto \sqrt{t}$ shortly thereafter.
  
\begin{figure}[t] 
\centering{
\includegraphics[scale=0.41]{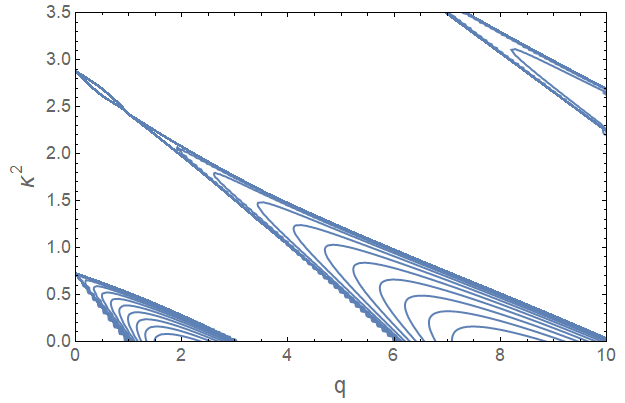}
}
\caption{ \label{Lame-fig}
Stability chart of the Lam\'e  equation \cite{Greene:1997fu} with $q=  \lambda_{\phi h} / (2 \lambda_\phi)$. The solution grows exponentially in the  areas enclosed by the contours. The Floquet exponent is constant along the contours and decreases outwards.
}
\end{figure}

The quantum Higgs field $\hat h$ can be expressed in terms of the creation and annihilation operators  $\hat a_k^\dagger ,\hat a_k$ as
\begin{equation}
\hat h = {1\over (2 \pi)^{3/2}} \int d^3 k \left(  \hat a_k h_k(t) e^{-i {\bf k\cdot x}}  + \hat a_k^\dagger   h_k(t)^* e^{i {\bf k\cdot x}}  \right) \;,
\label{mode-expansion}
\end{equation}
where $h_k$ are the Fourier modes with comoving 3--momentum ${\bf k} = {\bf p}/a$. 
 The equations of motion for the mode functions simplify in terms of the rescaled variables $X_k(t) \equiv a(t) h_k(t) $,
   \begin{equation}
X_k^{\prime\prime} + \left[     \kappa^2  + {\lambda_{\phi h} \over 2 \lambda_\phi} \;  {\rm cn}^2 \left( x, {1\over \sqrt{2}} \right)  \right] \,X_k=0 \;,
 \label{Lame}
 \end{equation}
  where the prime denotes differentiation with respect to $x$ and
     \begin{equation}
    \kappa^2  = {k^2 \over \lambda_\phi \Phi_0^2} \;.
     \end{equation}
This is known as the Lam\'e equation. Since the Jacobi cosine is a periodic function, the Floquet theory applies and the solutions either grow exponentially or oscillate in $x$,
depending  on $\kappa^2$ and $q = \lambda_{\phi h} / (2 \lambda_\phi)$. The corresponding stability chart is shown in Fig.\,\ref{Lame-fig}.
In the ``unstable'' regions, the amplitude grows exponentially representing {\it parametric resonance} and leading to Higgs production. The Higgs  $k$--mode occupation numbers are then found via
\begin{equation}
n_k = {\omega_k \over 2} \, \left(  {   \vert \dot X_k \vert^2   \over \omega_k^2} + \vert  X_k \vert^2   \right) -{1\over 2}    \;,
\end{equation}
where $ \omega_k^2 = \kappa^2  + {\lambda_{\phi h} \over 2 \lambda_\phi} \;  {\rm cn}^2 \left( x, {1\over \sqrt{2}} \right) $.
The corresponding  $comoving$ variance and energy density (in the Hartree approximation) are given by
\begin{equation}
\langle X^2 \rangle = \int { d^3 k \over (2\pi)^3}\; \vert X_k \vert^2  \simeq \int { d^3 k \over (2\pi)^3}\; {n_k \over \omega_k}~~~,~~~
\rho_X =  \int { d^3 k \over (2\pi)^3}\; \omega_k n_k \;.
\end{equation}

Note that the behaviour of the solution is determined by the $ratio$ of the couplings, $ \lambda_{\phi h} / (2 \lambda_\phi)$. Therefore, even for small couplings the  
resonance can be strong, especially if $ \lambda_{\phi h} / (2 \lambda_\phi) \gg 1$. Since  the system is  conformal, the resonance does not stop due to the Universe expansion. It only terminates due to backreaction of the produced particles and rescattering. For $ \lambda_{\phi h} / (2 \lambda_\phi) \ll 1$, the resonance becomes narrow, with a suppressed Floquet exponent, 
and the exponential growth of the amplitude is only seen  on a large timescale.

 Due to  self--interaction, an oscillating inflaton background also generates inflaton quanta. Indeed,   expanding $\phi = \langle \phi  \rangle + \delta \phi$ and quantizing $\delta \phi$, one can write down 
 an equation of motion for the inflaton  analog of $X_k$, which we denote by $\varphi_k$. Since at quadratic order $\phi^4 \rightarrow 6 \langle \phi  \rangle^2 \, \delta \phi^2$, we find
 \begin{equation}
\varphi_k^{\prime\prime} + \left[     \kappa^2  + 3 \;  {\rm cn}^2 \left( x, {1\over \sqrt{2}} \right)  \right] \,\varphi_k=0 \;.
 \label{Lame1}
 \end{equation}
 Therefore, the inflaton quanta are generated with the effective $q$--parameter $ \lambda_{\phi h} / (2 \lambda_\phi)  \rightarrow 3$.
 
 For $\lambda_{\phi h} / (2 \lambda_\phi)  \ll 3$, the inflaton quanta production is more efficient than Higgs production. In this case, most of the energy remains in the dark sector and 
 no successful reheating occurs. Therefore, in what follows we exclude this possibility and focus on moderate and large $\lambda_{\phi h} / (2 \lambda_\phi)$.
 
 In reality, particle production is complicated by backreaction and rescattering of the produced particles \cite{Khlebnikov:1996mc},\cite{Prokopec:1996rr}, which is not captured by the Lam\'e equation. Therefore, lattice simulations
 are often necessary for understanding the dynamics of the system.
 
\subsection{Lattice simulations}

In the coupling range of interest, $\lambda_{\phi h} \gtrsim \lambda_\phi$, the parametric resonance is sufficiently strong such that the perturbative estimates
are inadequate.  In this regime, 
the occupation numbers are large which enables the use of classical   lattice simulations. These are essential for capturing the backreaction and rescattering effects, especially 
when the dynamics become highly non--linear. Related lattice studies and tools have recently appeared in \cite{Figueroa:2016wxr},\cite{Figueroa:2021yhd}.

Our main goal is to understand how much energy can be transferred from the inflaton to the   Higgs field, which determines the composition of the Universe.
In a realistic scenario, almost all of the inflaton energy must be converted in the SM radiation.
Since the efficiency of particle production depends on the ratio of the couplings $\lambda_{\phi h} / \lambda_\phi$, in our simulations we set $\lambda_\phi = 10^{-13}$ and vary $\lambda_{\phi h}$.
The allowed values of $ \lambda_{\phi h}$ are bounded from above for non--thermal dark matter. First, DM should not thermalize and, second, the Higgs thermal bath should
 not produce too much DM via freeze--in \cite{Yaguna:2011qn}, which is  operative for reheating temperatures above 20 GeV or so. 
 The second constraint is stronger: it requires $\lambda_{\phi h} < 2 \times 10^{-11}$ for $m_\phi \gtrsim m_h$ and $\lambda_{\phi h} <  10^{-11}\; \sqrt{ {\rm GeV} /m_\phi}$ for $m_\phi \ll m_h$ \cite{Lebedev:2019ton}.
 This implies
 \begin{equation}
 \lambda_{\phi h } \lesssim 10^{-9}
 \end{equation}
  as long as dark matter is warm or cold, $m_\phi 
 \gtrsim 10$ keV, as required by   structure formation. If the reheating temperature is very low, the bound gets weaker, although this does not affect 
 our analysis of particle production.
 
 The realistic system is very complicated: the Higgs production is accompanied by its decay  into other SM states and their thermalization. To account for all of the effects is a (currently) unsurmountable task, hence we have to resort to simplifications. We will consider the limiting cases, where either Higgs decay or Higgs production dominates. The result also depends on the Higgs self--coupling
 as it can induce significant backreaction. The value of the coupling at high energies is uncertain due to the uncertainty in the top quark mass, hence we take two representative values
 $\lambda_h =0$ and $\lambda_h =0.01$.
 
  The simulations are performed in conformal time $z$ defined by $dz =\sqrt{\lambda_\phi} \,\Phi_0  \,dt/a(t) $. It is equivalent to the variable $x$ at late times, which in practice means after a few oscillations. 
  At early times,  this relation is integrated numerically.

\subsubsection{Fast Higgs decay: no resonance} 
\label{fast-decay}
 
 If the Higgs quanta decay faster than they are generated, no resonant production takes place. 
 Since $\langle h \rangle \ll \sqrt{\lambda_{\phi h} } \phi $, the quarks are effectively massless compared to the Higgs and 
 the main decay channel is $h \rightarrow \bar t t$. The corresponding decay width is 
 \begin{equation}
 \Gamma_h = {3 y_t^2 \over 16 \pi} m_h^{\rm eff} \;,
 \end{equation}
 where $m_h^{\rm eff} = {\sqrt{\lambda_{\phi h} /2}} \, \phi$. Accounting for the time--dependence of $ \Gamma_h$,
 this results in the particle number decrease as $\exp (-2\Gamma_h \,t)$. On the other hand, the $k$--mode occupation number during the resonance grows as $\exp (2\mu_k z)$,
 where $\mu_k$ is the Floquet exponent. Using $z= (48 \lambda_\phi)^{1/4} \sqrt{t}$ and $\phi \sim (3/\lambda_\phi)^{1/4}\, t^{-1/2}$ at late times, we find that the decay dominates for
  \begin{equation}
\lambda_{\phi h} \gtrsim 10^3 \lambda_\phi \;,
 \end{equation}
 with a typical value $\mu_k \sim 10^{-1}$. For smaller $\mu_k$, the bound decreases as $\mu_k^2$.
 Although this argument neglects possible backreaction of the produced quarks and effects of thermalization,
 it is clear that,  
 at very large $\lambda_{\phi h}$, the Higgs decay is important making   resonant Higgs production  inefficient.

     \begin{figure}[t] 
\centering{
\includegraphics[scale=0.56]{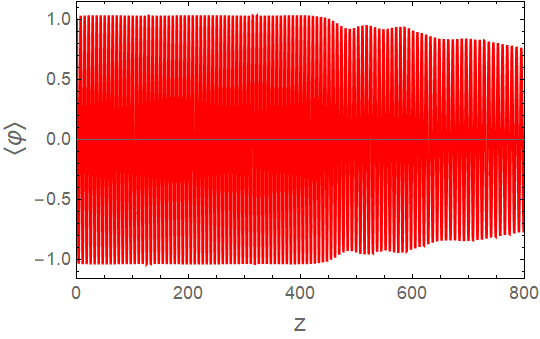}
}
\caption{ \label{zero-mode-decay}
Decay of the classical inflaton background  
$\langle \varphi \rangle $  in the comoving frame  ($ \varphi = a\, \phi$) due to emission of the inflaton quanta and rescattering. The conformal
time $z$ is defined by $dz =\sqrt{\lambda_\phi} \,\Phi_0  \,dt/a(t) $; $\lambda_\phi=10^{-13}$ and  $\Phi_0 =1.7$ in Planck units.
 The amplitude is normalized to 1 at the initial point. The simulation is performed with LATTICEEASY \cite{Felder:2000hq}.
}
\end{figure}

On the other hand,
 the resonant production of the inflaton quanta proceeds unimpeded according to Eq.\,\ref{Lame1}. Simulations show that the resonance is terminated by backreaction at \cite{Khlebnikov:1996mc}
 \begin{equation}
 z_* = 76-14.3 \ln \lambda_\phi \;.
 \end{equation} 
 After that, the inflaton zero mode decays due to rescattering. Specifically, defining the comoving inflaton background  as $\varphi = a \phi$,
 one finds $\varphi \propto z^{-1/3}$. Since the energy density of the zero mode scales as $\varphi^4$,  
 within conformal time $2z_*$ most of the energy is contained in the $fluctuations$. An example of the zero mode decay  for $\lambda_\phi =10^{-13}$ obtained with LATTICEEASY  \cite{Felder:2000hq} simulations
 is shown in Fig.\,\ref{zero-mode-decay}. This simulation includes the inflaton field only.
 We find that  the resonance ends around $z_*\simeq 500$ and most of the energy gets converted into inflaton fluctuations by $z\sim 800$.

The perturbative Higgs production is highly suppressed in this regime. The Higgs pairs are generated by an oscillating classical background \cite{Dolgov:1989us},\cite{Ichikawa:2008ne}, 
and the corresponding inflaton
decay rate for  4  $massless$ Higgs d.o.f. is given by \cite{Lebedev:2021xey}
\begin{equation}
\Gamma_\phi^{\rm pert} = C \, {\lambda_{\phi h}^2 \over 16\pi  }\, {\Phi_0 \over a(t)  \sqrt{\lambda_{\phi}} }\;,
\end{equation}
 where $C \simeq 0.4$. Requiring $\Gamma_\phi^{\rm pert} \, t \sim 1$ translates into the conformal decay time of order $z\sim 10^2 \lambda_\phi / \lambda_{\phi h}^2$. 
 This is already much longer than the characteristic decay time into inflaton fluctuations, $z \sim 10^3$. In reality, the Higgs production is much more suppressed since the 
 Higgs is heavy: the effective Higgs mass is much greater than the principal oscillation frequency of the inflaton, $  m_h^{\rm eff} = {\sqrt{\lambda_{\phi h} /2}} \, \phi \gg \omega \sim \sqrt{\lambda_\phi} \, \phi$.
 This leads to exponential suppression of the decay rate by $e^{-2\pi\, m_h^{\rm eff} /\omega}$ \cite{Lebedev:2021xey}. Thus, perturbative decay can be neglected.

We conclude that the inflaton background decays primarily into inflaton fluctuations.  Feebleness of their interactions ensures that the reaction rates are slower than the Hubble 
expansion\footnote{Relevant thermalization constraints have been considered carefully in \cite{DeRomeri:2020wng},\cite{Arcadi:2019oxh}.}
so that the inflaton quanta
   get never converted into SM radiation and 
\begin{equation}
\rho_\phi \gg \rho_{\rm SM} \;.
\end{equation}
 These quantities scale as radiation until the inflaton becomes non--relativistic, which makes the inequality even stronger. 
 The resulting Universe is therefore dark and unrealistic.

\subsubsection{Slow Higgs decay, $\lambda_h =0$}
\label{SlowDecay-0}

\begin{figure}[t] 
\centering{
\includegraphics[scale=0.37]{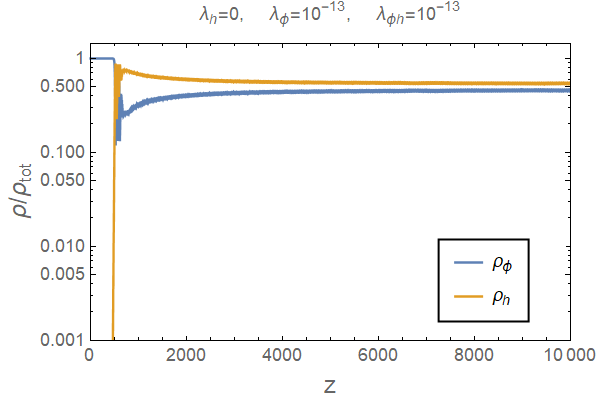}
\includegraphics[scale=0.37]{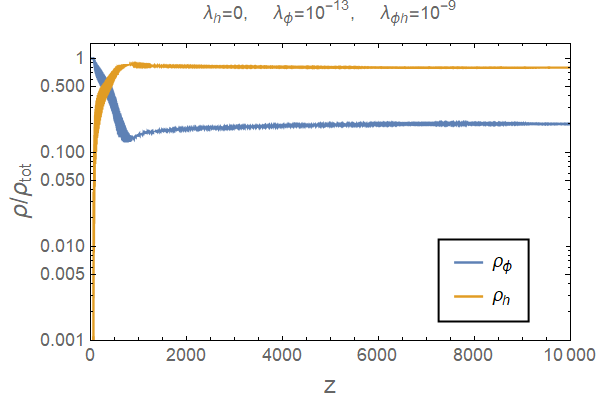}\\
\vspace{0.3cm}
\includegraphics[scale=0.41]{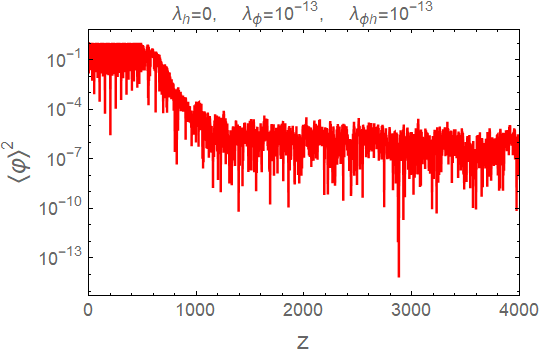}
\includegraphics[scale=0.41]{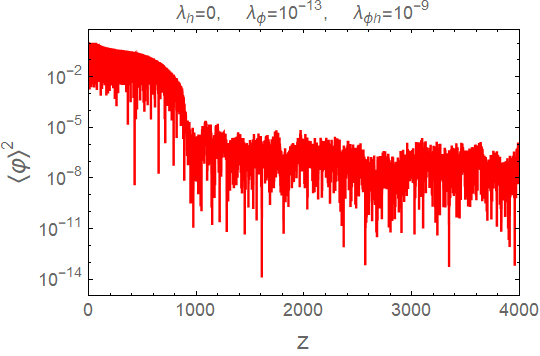}
}
\caption{ \label{rho-fig}
{\it Upper row:} fraction of the energy density carried by 4 Higgs  d.o.f. and the inflaton as a function of the conformal time $z$,  
  $dz =\sqrt{\lambda_\phi} \,\Phi_0  \,dt/a(t) $ and  $\Phi_0 =1.7$ in Planck units, obtained with LATTICEEASY. 
  {\it Lower row:} decay of the classical inflaton background  $\langle \varphi \rangle^2 $ in the comoving frame 
  ($ \varphi = a\, \phi$). The amplitude is normalized to 1 at the initial point.}
\end{figure}

Consider now another idealized situation: 
suppose the Higgs decay can be neglected on the timescale of the resonance and rescattering,  i.e. when $\lambda_{\phi h}$ is not too large or Higgs decay is impeded by backreaction.   
For $ \lambda_{\phi h} \gtrsim \lambda_\phi$, the resonance is strong and results in explosive Higgs production.  As shown in Fig.\,\ref{rho-fig},
 a large fraction of the initial inflaton energy density can be converted into the Higgs quanta. 
 Here we include all 4 Higgs degrees of freedom $h_i$ available at high energies. We observe that the inflaton zero mode decays quickly and converts into
 fluctuations.

At large couplings, $\lambda_{\phi h} \sim 10^{-10}$, the system tends to quasi--equilibrium on a relatively short time scale, $z\sim 10^3$. Although the distribution is not yet Bose--Einstein, 
the energy is distributed almost equally  among the available degrees of freedom. For $\lambda_{\phi h} \gtrsim 10^{-10}$, roughly 80\%
of the energy is carried by the Higgs quanta and 20\% by the inflaton, while for smaller couplings the inflaton fraction is higher (Fig.\,\ref{rho-fig}).
We have verified that with a single Higgs d.o.f., the maximal Higgs fraction is about 50\%. 

It should be noted that,  according to the previous subsection, $\lambda_{\phi h}$ values above $10^{-10}$ are expected to lead to fast Higgs decay. Nevertheless, we still 
consider such large couplings in this subsection since the Higgs decay may be blocked by backreaction, e.g. thermal masses of the decay products.  
Such  effects are hard to evaluate precisely at this stage given the multitude of possible processes and timescales, hence one should not  exclude the range $\lambda_{\phi h} > 10^{-10}$
from consideration.

In a  realistic situation, we expect the Higgs field to channel at least some of its energy 
 into other SM states. If this process is fast enough, the system may reach quasi--equilibrium where
  the energy would be  shared almost democratically by the relativistic degrees of freedom,
 \begin{equation}
 {\rho_\phi \over \rho_{\rm tot}} \sim {1\over \# \, {\rm d.o.f.}}
 \end{equation}
When rescattering stops, the total number of the inflaton quanta remains approximately constant.
This allows us to obtain the $lower$ bound on the dark matter abundance $Y$, which also remains invariant after this stage.
 It is defined by $Y=n_\phi/s_{\rm SM}$, where  $n_\phi$ is the inflaton number density and  $s_{\rm SM}$
is the entropy density of the SM fields.
For a thermalised SM sector  in the relativistic regime, 	$s_{\rm SM}$  is close the SM quanta number density, up to a factor of a few: $s_{\rm SM} \sim 4 n_{\rm SM}$. 
 Since
$n_\phi /n_{\rm SM} \sim \rho_\phi / \rho_{\rm SM}$, 
  for 
$107$ SM degrees of freedom at high temperature, we find 
\begin{equation}
Y \gtrsim 10^{-3} \;.
\label{Y}
\end{equation}
 This number is far above the observed value $Y_{\rm obs} = 4 \times 10^{-10} \; {\rm GeV}/m_\phi $, given that $m_\phi \gtrsim
10$ keV to be consistent with the structure formation constraints. 
Thus, the  emerging Universe is again unacceptably dark.

\subsubsection{Slow Higgs decay, $\lambda_h =0.01$}

\begin{figure}[t!] 
\centering{
\includegraphics[scale=0.37]{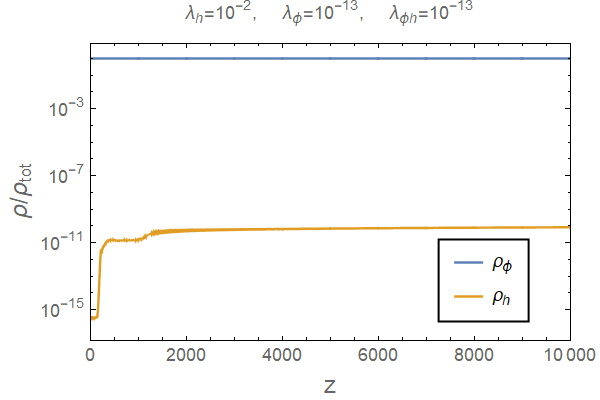}
\includegraphics[scale=0.37]{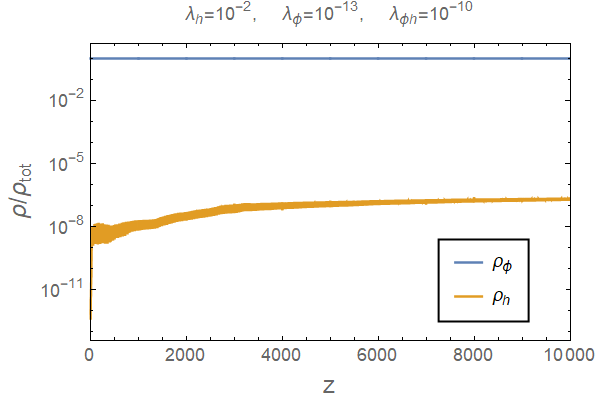}
}
\caption{ \label{rho-fig-1}
Fraction of the energy density carried by 4 Higgs  d.o.f. with self--interaction and the inflaton as a function of the  conformal time $z$, 
  obtained with LATTICEEASY. $\Phi_0=1.7$ in Planck units.
}
\end{figure}

The Higgs self--interaction has a profound effect on the dynamics of the system \cite{Prokopec:1996rr}. A significant self--coupling induces a large effective mass term shutting down the resonance.
This occurs when the Higgs fluctuations, characterized by the variance  $\langle h^2 \rangle$, grow large such that 
 the effective inflaton and Higgs masses become comparable,
\begin{equation}
\lambda_\phi \phi^2 \sim \lambda_h \langle h^2 \rangle \;.
\end{equation}
 After that, the explosive growth of the Higgs amplitude stops. Subsequently,  
 $\langle h^2 \rangle$ evolves non--linearly: it 
 decreases and increases again before stabilizing eventually. As shown in Fig.\,\ref{rho-fig-1}, only a tiny fraction of the inflaton energy can be converted into the Higgses 
 for a realistic self--coupling.\footnote{This ceases to be true at very large $\lambda_{\phi h} \gtrsim \lambda_h$, which however leads to thermal dark matter.}
 Since $\phi^2 $ and $\langle h^2 \rangle$ redshift the same way, this remains true at later times as well.

 Although the Higgs production is hindered by backreaction, the background decay into the inflaton quanta continues. We thus end up with a situation similar to that discussed in  
 Section\,\ref{fast-decay}, 
 \begin{equation}
\rho_\phi \gg \rho_{\rm SM} \;.
\end{equation}
 If one were to account for Higgs decay which reduces   $\langle h^2 \rangle$, the system would interpolate between those of Section\,\ref{fast-decay} and
 Section\,\ref{SlowDecay-0}. Either way, the resulting Universe is unrealistic.

We conclude that backreaction and rescattering make a crucial impact on the reheating process in the framework of inflaton dark matter,
rendering the non--thermal dark matter option unrealistic.   
We find similar results for a $\phi^2$ inflaton potential: for a small Higgs portal coupling, the energy transfer to the SM radiation is suppressed,
while for a large coupling, the system reaches quasi--equilibrium and the inflaton retains too large a fraction of the total energy.

\section{Thermal inflaton dark matter}

If $\lambda_{\phi h}$ is sufficiently large, the system reaches thermal equilibrium via processes like $\phi h \rightarrow \phi h, \phi \phi \rightarrow hh$, etc.
Its  precursors are already seen in Fig.\,\ref{rho-fig}, right panels.
 The dark matter abundance is then determined  by the temperature and the usual freeze--out approach applies \cite{Silveira:1985rk},\cite{McDonald:1993ex},\cite{Burgess:2000yq}.
 The correct relic abundance can be obtained for parameter values allowing for efficient dark matter annihilation, subject to direct DM detection and perturbativity constraints.

Away from the narrow resonance region $m_\phi \simeq 62$ GeV, efficient DM annihilation $\phi \phi \rightarrow \;$SM combined with the XENON1T   bound requires \cite{Athron:2018ipf}
\begin{equation}
\lambda_{\phi h} (1\,{\rm TeV}) \gtrsim 0.25 \;.
\end{equation}
This bound applies at the TeV scale, while the couplings at the inflation scale are obtained by the solving the renormalization group (RG) equations. The list of the RG equations can be found in \cite{Lebedev:2021xey}, while
the most important one for us reads 
\begin{equation}
16 \pi^2 {d \lambda_\phi \over d t} = 2 \lambda_{\phi h}^2 + 18 \lambda_\phi^2 \;,
\end{equation}
where $t = \ln\,\mu$ with $\mu$ being the RG energy scale. This implies, in particular, that the inflaton self--coupling at least of the size  $\lambda_{\phi h}^2 /(8\pi^2)$ gets generated (ignoring 
a large $log$),
\begin{equation}
\lambda_{\phi } (H) \gtrsim 10^{-3} \;.
\end{equation}
In other words, the Higgs--inflaton coupling induces the Coleman--Weinberg contribution to the (Jordan frame) inflaton potential, 
  \begin{equation}
\Delta V_{\rm 1-loop}\simeq {\lambda_{\phi h}^2 \over 64 \pi^2} \, \phi^4 \ln { \phi^2 \over \phi_*^2} \;,
\end{equation} 
 where  $\phi_*$ is a reference field value and  4 Higgs degrees of freedom have been included. By choosing $\phi_*$ appropriately, one can suppress the correction
 at one point, but not over the entire field range where the last 60 e--folds of inflation take place.
Clearly, the generated coupling $\lambda_\phi$ is far above the unitarity bound (\ref{uni-bound}), which signals inconsistency of the model.

The above conclusion is evaded  at the Higgs resonance, 
\begin{equation}
m_\phi \simeq m_{h_0}/2 \;,
\end{equation}
where $m_{h_0} =125$ GeV is the physical Higgs mass.
In this case, resonant DM annihilation $\phi\phi \rightarrow h \rightarrow {\rm SM}$
 is efficient even for small couplings $\lambda_{\phi h} \gtrsim 10^{-4}$ \cite{Athron:2018ipf}, although possible complications related to  early kinetic decoupling should be kept in mind
 \cite{Binder:2017rgn}. 
 The resulting correction to the inflaton self--coupling is negligible and all of the constraints are satisfied. 
 We note however that $| m_\phi - m_h/2 |$ must be below a few GeV, which appears rather unnatural yet not impossible.

\section{Extensions}

Our results depend critically on the minimality assumption, which serves as the main motivation for inflaton dark matter. 
In particular, we require the minimal number of degrees of freedom consistent with observations.
Nevertheless,
it is worthwhile to explore further options.

The unitarity problem  can be evaded in the Palatini formalism, i.e. at the price of adding  extra (gravitational) degrees of freedom \cite{Bauer:2008zj}. In this case, the connection $\Gamma_{\alpha\beta}^\lambda$ 
is promoted to a dynamical variable along with the metric. One finds that the unitarity cutoff of this theory is of order (in Planck units) \cite{Bauer:2010jg}
\begin{equation}
\Lambda_{\rm Pal} \sim {1\over \sqrt{\xi_\phi} }\;,
\end{equation}
which is above the inflationary scale. As a result, a large inflaton self--coupling
can be consistent with unitarity, thus opening  the possibility of a TeV scale thermal inflaton DM. 
A somewhat uncomfortable aspect of this model is that the requisite $\xi_\phi$ has to be very large,  of order  $10^7$ or above.

 In our reheating analysis, we have set  
the non--minimal Higgs coupling to gravity $\xi_h$ to zero.  Nevertheless, its non--zero value 
subject to  (\ref{constraint-xi}) is not expected to affect the results in any significant way,
even if it makes the Higgs production more efficient \cite{Ema:2017loe},\cite{Fairbairn:2018bsw}. 
Indeed, the main issue we find is that  efficient particle production leads to quasi--equilibrium, where the energy is distributed almost equally
among the constituents. Thus, the inflaton carries at least  1\% of the energy of the system. At weak coupling and for realistic Higgs self--interaction,
the resonance is inefficient and  this fraction is much larger.
In both cases, dark matter is  overabundant unless it is capable of annihilating  efficiently.

One may also relax the assumption that $\phi$ alone drives inflation. In general, the inflaton  may be a combination of the Higgs and the DM field $\phi$.  
At large field values,  there exists a stable flat direction \cite{Lebedev:2011aq}
\begin{equation}
{h \over \phi}= \sqrt{  2 \lambda_\phi \xi_h -       \lambda_{\phi h } \xi_\phi  
 \over   2 \lambda_h \xi_\phi -       \lambda_{\phi h } \xi_h  }
\end{equation}
if both the numerator and denominator under the square root are positive, and $\xi_\phi + \xi_h \gg 1$.  Slow roll along this direction would lead
to inflation with predictions close to those of Eq.\,\ref{n-r}.
The efficiency of reheating can then be increased if the Higgs proportion is large, $h/\phi \gg 1$.
In that case, however, 
inflation is driven mostly by the Higgs such that 
one faces the usual Higgs inflation unitarity problem.\footnote{A mixed Higg--singlet inflaton scenario was explored in \cite{Ballesteros:2016euj}. The above conclusion does not apply to this model since the singlet is unstable by construction. }

\section{Conclusion}

The concept of inflaton dark matter is interesting in that it is economical: a single field is responsible for both inflation
and dark matter. We have considered the minimal framework which fits the inflationary data very well and where inflation
is driven by a non--minimal scalar coupling to gravity. The parity symmetry guarantees that the inflaton remains as a stable relic
and can potentially play the role of dark matter. 

The focus of this work is on understanding the reheating processes in such a 
framework. We have examined both the thermal and non--thermal dark matter options.
One of the important features of the system is that the inflaton background decays non--perturbatively,
thus necessitating lattice simulations to describe. We find that,
for non--thermal DM, the reheating is hindered by backreaction and rescattering effects 
resulting in overabundance of dark matter.  
For thermal dark matter, radiative corrections to the inflaton potential play an important role. 
In this case, the correct relic abundance can be produced, yet  
the unitarity constraint forces the inflaton mass into a very narrow window
close to half the Higgs mass.

 These conclusions can be evaded in less minimalistic set--ups which invoke additional degrees of freedom.
\\ \ \\
{\bf Acknowledgements.} We are thankful to Yohei Ema for helpful discussions and sharing some of his results.

 \appendix

\section{Simulation details}

In this work, we have performed lattice simulations with CLUSTEREASY, the parallel computing version of LATTICEEASY. 
For most purposes, the dimension of the lattice was set to $D=3$ and the number of the grid points per edge was fixed at $N=128$ ($128^3$ in total).
The simulations mainly target the late time behaviour of the system, for which the UV momentum spectrum is essential.  To capture the relevant features, 
we have made the upper bound of the momentum space $k_{max}$ (in LATTICEEASY convention) dynamical, i.e. $\lambda_{\phi h}$--dependent.
The size of the box $L$ in rescaled distance units was set to  
\begin{equation}
L=\frac{\pi \sqrt{D}N}{40} \left( \frac{ \lambda_{\phi h}}{\lambda_{\phi} (4 \pi^2)} \right)^{-0.25} \;,
\end{equation}
such that
\begin{equation}
k_{max}=k_{min} \times  \frac{\sqrt{D}}{2}N = \frac{2\pi}{L}  \frac{\sqrt{D}}{2}N=40 \times  \left( \frac{ \lambda_{\phi h}}{\lambda_{\phi} (4 \pi^2)} \right)^{0.25} \;,
\end{equation}
where $k_{min}$ represents the lower bound of the momentum space in rescaled units and the prefactor 40 has been  determined empirically.
To verify  reliability of our results, we have run extended 2D simulations with  $N=1024$ which also capture the relevant IR physics. We find that 
the late time distributions are indeed 
consistent.

\end{document}